\documentclass[aps,prl,twocolumn,groupedaddress,showpacs]{revtex4}

\usepackage{bm}
\usepackage{graphicx}

\begin{document}

\title{Universality of superconducting gaps in overdoped Ba$_{0.3}$K$_{0.7}$Fe$_2$As$_2$ observed by angle-resolved photoemission spectroscopy}

\author{
	K. Nakayama,$^{1,2}$
	T. Sato,$^{1,3}$
	P. Richard,$^{4,5}$
	Y.-M. Xu,$^6$
	T. Kawahara,$^1$
	K. Umezawa,$^1$
	T. Qian,$^5$
	M. Neupane,$^6$
	G. F. Chen,$^7$
	H. Ding,$^5$
	and T. Takahashi$^{1,2,3}$}

\affiliation{$^1$Department of Physics, Tohoku University, Sendai 980-8578, Japan}
\affiliation{$^2$CREST, Japan Science and Technology Agency (JST), Kawaguchi 332-0012, Japan}
\affiliation{$^3$TRiP, Japan Science and Technology Agency (JST), Kawaguchi 332-0012, Japan}
\affiliation{$^4$WPI Research Center, Advanced Institute for Materials Research, Tohoku University, Sendai 980-8577, Japan}
\affiliation{$^5$Beijing National Laboratory for Condensed Matter Physics, and Institute of Physics, Chinese Academy of Sciences, Beijing 100190, China}
\affiliation{$^6$Department of Physics, Boston College, Chestnut Hill, MA 02467, USA}
\affiliation{$^7$Department of Physics, Renmin University of China, Beijing 100872, China}

\date{\today}

\begin{abstract}
We have performed angle-resolved photoemission spectroscopy on the overdoped Ba$_{0.3}$K$_{0.7}$Fe$_2$As$_2$ superconductor ($T_c$ = 22 K). We demonstrate that the superconducting (SC) gap on each Fermi surface (FS) is nearly isotropic whereas the gap value varies from 4.4 to 7.9 meV on different FSs. By comparing with under- and optimally-doped Ba$_{1-x}$K$_x$Fe$_2$As$_2$, we find that the gap value on each FS nearly scales with $T_c$ over a wide doping range (0.25 $\leq$ $x$ $\leq$ 0.7). Although the FS volume and the SC gap magnitude are strongly doping dependent, the multiple nodeless gaps can be commonly fitted by a single gap function assuming pairing up to the second-nearest-neighbor, suggesting the universality of the short-range pairing states with the $s_{\pm}$-wave symmetry.
\end{abstract}


\maketitle

The discovery of Fe-based high-$T_c$ superconductors \cite{Kamihara} has generated a great interest on its superconducting (SC) mechanism. Angle-resolved photoemission spectroscopy (ARPES) has yielded an important insight into the pairing mechanism by directly observing the Fermi-surface (FS) dependent nodeless SC gaps on optimally doped (OPD) Ba$_{1-x}$K$_x$Fe$_2$As$_2$ \cite{Hong, Zhou, Hasan1, Borisenko, Nakayama}. However, in contrast to the relatively well-investigated SC gap properties in the OPD region, much less information is known about the gap properties in the underdoped (UD) \cite{Yiming1} and even less in the overdoped (OD) region. In particular, the experimental investigation of the SC gap properties in the latter region is crucial to elucidate the following fundamentally important issues: (1) the doping dependence of the pairing strength as well as its correlation with the $T_c$ value, which would shed light on the nature of the pairing force and the characteristics of the phase diagram. (2) the doping evolution of the pairing symmetry. So far, various competing gap symmetries, such as nodeless isotropic / anisotropic $s_{\pm}$-wave (extended $s$-wave with a sign change of the gap between different FSs), nodal $s_{\pm}$-wave, and nodal $d_{x^2-y^2}$-wave, have been theoretically predicted \cite{Mazin, Kuroki1, Wang, Seo, Yao, Cvetkovic, Kuroki2, Graser, Ikeda}, and possible transition of the pairing symmetry upon doping has been also suggested \cite{Kuroki2, Graser, Ikeda}. (3) the microscopic mechanism that determines the SC gap function. Although several gap functions have been proposed to describe the SC state \cite{Mazin, Kuroki1, Wang, Seo, Yao, Cvetkovic, Kuroki2, Graser, Ikeda}, the applicability of these models is currently under intensive debate. These unresolved issues demand direct experiments to elucidate the doping evolution of the momentum ($k$) and the FS dependences of the SC gap.

In this Rapid Communication, we report high-resolution ARPES results on OD Ba$_{0.3}$K$_{0.7}$Fe$_2$As$_2$ ($T_c$ = 22 K). We precisely determine the FS topology, the SC gap value and its $k$ dependence, and compare them with the results obtained on the UD and OPD Ba$_{1-x}$K$_x$Fe$_2$As$_2$. We demonstrate that the SC gap size in this OD region is nearly isotropic on each FS but strongly FS dependent. The SC gap values linearly scale with $T_c$ over a wide doping range. We discuss the implications of these observations as well as the universal features among UD, OPD and OD samples.

High-quality single crystals of OD Ba$_{0.3}$K$_{0.7}$Fe$_2$As$_2$ ($T_c$ = 22 K; labeled as K0.7) were grown by the FeAs-flux method \cite{GFChen1}. OPD ($T_c$ = 37 K; K0.4) and UD ($T_c$ = 26 K; K0.25) samples \cite{GFChen2} were also studied for comparison. ARPES measurements were performed at Tohoku University using a SES2002 spectrometer with a He discharge lamp ($h\nu$ = 21.218 eV). The energy and angular resolutions were set at 4-12 meV and 0.2$^{\circ}$, respectively. Clean surfaces for the ARPES measurements were obtained by cleaving crystals $in$ $situ$ in a working vacuum better than 5$\times$10$^{-11}$ Torr. The Fermi energy ($E_F$) of the samples was referenced to that of a gold film evaporated onto the sample holder.

\begin{figure*}[!t]
\begin{center}
\includegraphics[width=6in]{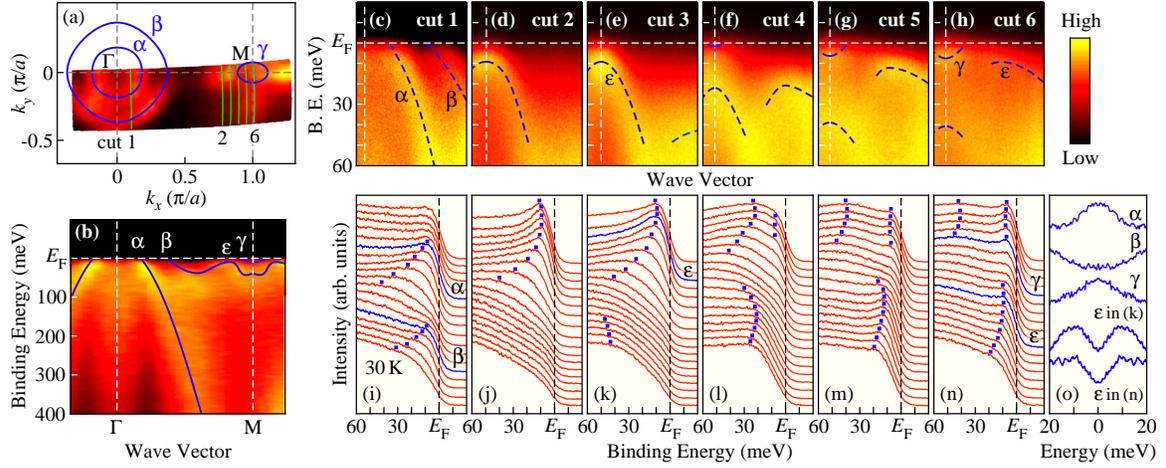}
\end{center}
\caption{
(color online) (a) Plot of ARPES intensity at $E_{\rm F}$ as a function of the in-plane wave vector measured at 10 K for the K0.7 sample (OD; $T_c$ = 22 K). The intensity is obtained by integrating the spectral intensity within $\pm$10 meV with respect to $E_{\rm F}$. Solid curves show the schematic FSs. (b) ARPES intensity plot along the $\Gamma$-M line as a function of binding energy and wave vector. Solid curves are guides for the eyes to trace the band dispersion. (c)-(h) ARPES intensity plots in the vicinity of $E_{\rm F}$ and (i)-(n) corresponding EDCs measured in the normal state (30 K) along cuts 1-6 indicated by green lines in (a). Dashed curves and dots are guides for the eyes. White vertical dashed lines in (c)-(h) represent the $\Gamma$-M line. (o) Symmetrized EDCs of the blue (dark gray) curves in (i), (k) and (n).
}
\end{figure*}

Figure 1(a) shows the ARPES intensity at $E_{\rm F}$ plotted as a function of the in-plane wave vector for the K0.7 sample. To accurately determine the FS topology, we measured normal-state band dispersion with an energy resolution as high as 4 meV [Figs. 1(c)-(n)]. As shown in Figs. 1(c) and (i), we identify two highly dispersive bands (called $\alpha$ and $\beta$) which create hole FSs centered at the $\Gamma$ point. Near the M point, which corresponds to the $(\pi, 0)$ point in the one-Fe unit cell description, we observe a shallow electron pocket ($\gamma$). The bottom of the $\gamma$ band is within 10 meV below $E_{\rm F}$ and thus creates a small electron FS centered at the M point. We also find a holelike band ($\epsilon$) which originates from the hybridization of the $\beta$ band with the electronlike band [Fig. 1(b)]. While the $\epsilon$ band crosses $E_{\rm F}$ in extremely OD KFe$_2$As$_2$ (labeled as K1) \cite{Sato}, the top of the $\epsilon$ band is still below $E_{\rm F}$ in this K0.7 sample. This is evident from the presence of two peaks in the symmetrized spectra [Fig. 1(o)] that are obtained at the $k$ location where the $\epsilon$ band shows a local maximum. The present results thus demonstrate that there are large hole and small electron FSs at the $\Gamma$ and M points in the K0.7 sample, respectively. We note that recent ARPES studies on the OPD sample reported the presence of a third holelike band crossing $E_F$ and its degeneracy with the $\alpha$ band at $k_z$ $\sim$ 0 \cite{Yiming2, Feng}.  Since the present result obtained by the He I line ($h\nu$ = 21.218 eV) basically reflects the electronic structure around the $k_z$ $\sim$ 0 plane \cite{Yiming2, Feng}, the observed $\alpha$ band would be composed of degenerated couple of bands. It is also important to note that the present result does not show the FS reconstruction observed in some of the previous studies on Fe-based superconductors \cite{Shen, Hasan2}. In addition, the observed FS volume consistently evolves while changing the bulk hole concentration [Figs. 4(d)-(f)]. These facts suggest the bulk origin of the observed electronic structure.

\begin{figure}[!t]
\begin{center}
\includegraphics[width=3in]{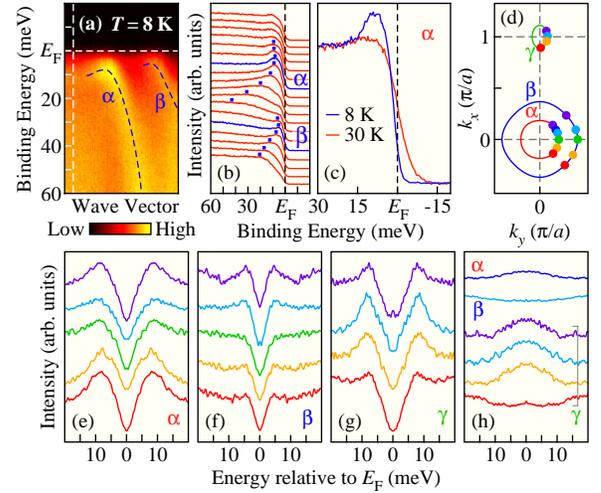}
\end{center}
\caption{
(color online) (a) ARPES intensity plot as a function of binding energy and wave vector measured at 8 K near the $\Gamma$ point and (b) corresponding EDCs. Dashed curves and dots are guides for the eyes to trace the dispersion of the $\alpha$ and $\beta$ bands.  EDCs at $k_{\rm F}$ points are indicated by blue (dark gray) curves in (b). (c) High-resolution EDCs near $E_{\rm F}$ at $T$ = 8 K and 30 K measured at a $k_{\rm F}$ point of the $\alpha$ band. (d) Schematic FSs of the K0.7 sample. (e)-(g) Symmetrized EDCs at 8 K measured at various $k_{\rm F}$ points of the $\alpha$, $\beta$ and $\gamma$ FSs indicated by filled circles in (d). The color of the spectra is the same as that of the circles. (h) Symmetrized EDCs at 30 K for representative $k_F$ points on each FS.
}
\end{figure}

Figures 2(a) and (b) show the ARPES intensity plot and the corresponding energy distribution curves (EDCs) measured in the SC state. Both the $\alpha$ and $\beta$ bands show a local energy maximum at their corresponding Fermi wave vectors ($k_{\rm F}$) and disperse back toward higher binding energy. This behavior is reminiscent of the dispersion relation of Bogoliubov quasiparticles in a superconductor, indicating the opening of a SC gap. The SC gap is more clearly seen in Fig. 2(c), where the 8 K EDC at $k_{\rm F}$ of the $\alpha$ band exhibits a leading edge shift toward higher binding energy and the emergence of a coherent peak as compared to the 30 K EDC in the normal state. To accurately determine the gap size and its $k$ dependence, we have performed high-resolution measurements ($\Delta$$E$ = 4 meV) at various $k_{\rm F}$ points of the $\alpha$, $\beta$ and $\gamma$ bands. The EDCs obtained after the symmetrization procedure are shown in Figs. 2(e)-(h). The symmetrized EDCs below $T_c$ [Figs. 2(e)-(g)] show two peaks irrespective of the $k_{\rm F}$ position, indicating the nodeless character of the SC gap at $k_z$ $\sim$ 0 in the K0.7 sample. The present result rules out the possibility of $d$-wave or extended $s$-wave superconductivity with vertical line nodes, but leaves open the possibility for having horizontal nodes away from $k_z$ = 0. It has been reported that gap nodes appear on the K1 compound \cite{Dong, Fukazawa, Hashimoto}, implying a possible transition from nodeless to nodal pairing states in the OD region. Our observation of nodeless gap in the K0.7 sample indicates that $x$ = 0.7 is a lower limit for the critical doping of the transition, if the nodes appear vertically. From the viewpoint of the electronic structure, the significant difference between the K1 and $x$ $\leq$ 0.7 samples is the FS topology near the M point where the shrunk electron FS in K0.7 completely vanishes in K1. It is thus inferred that the change of FS topology could be associated with the transition of the pairing symmetry. Based on the rigid band shift model, we suggest that the critical doping would be 0.8 $\leq$ $x$ $\leq$ 0.9.

\begin{figure}[!t]
\begin{center}
\includegraphics[width=3in]{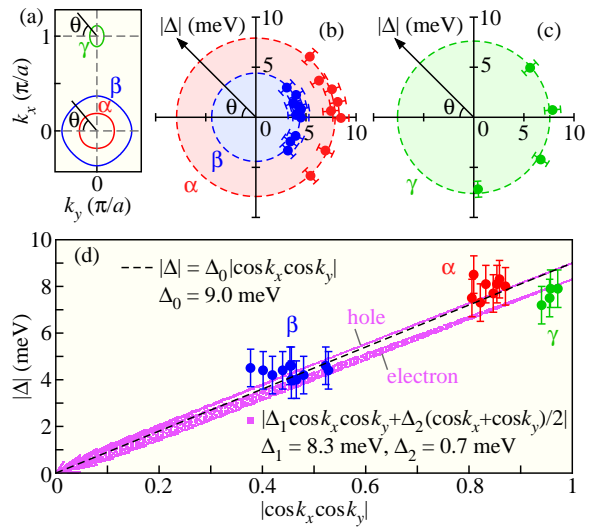}
\end{center}
\caption{
(color online) (a) Schematic FSs and definition of the FS angle ($\theta$). (b) and (c) Polar plots of the SC gap size at 8 K for the (b) $\alpha$, $\beta$ and (c) $\gamma$ FSs as a function of $\theta$. Dashed circles show the averaged SC gap values on each FS. (d) Plot of the SC gap size as a function of $\mid$$cosk_xcosk_y$$\mid$. The dashed line and purple dots represent the best fit assuming the gap function $\mid$$\Delta$$\mid$ = $\Delta_0$$\mid$$cosk_xcosk_y$$\mid$ and $\mid$$\Delta$$\mid$ = $\mid$$\Delta_1cosk_xcosk_y+\Delta_2(cosk_x+cosk_y)/2$$\mid$, respectively.
}
\end{figure}

To clarify the isotropy / anisotropy of the gap function in the K0.7 sample, we have estimated the SC gap size ($\mid$$\Delta$$\mid$) from the energy separation between the coherence peak and $E_{\rm F}$, and plotted the gap size as a function of the FS angle ($\theta$) in Figs. 3(b) and (c). The results confirm the absence of vertical line nodes in the K0.7 sample. It is also apparent that the gap is strongly FS dependent but nearly isotropic on each FS. We obtained averaged gap values $\mid$$\Delta$$\mid$ of 7.9 $\pm$ 0.8, 4.4 $\pm$ 0.8 and 7.6 $\pm$ 0.8 for the $\alpha$, $\beta$ and $\gamma$ FSs, respectively, corresponding to 2$\mid$$\Delta$$\mid$/$k_BT_c$ ratios of 8.3 $\pm$ 0.9, 4.6 $\pm$ 0.9 and 8.0 $\pm$ 0.9. These experimental results clearly show the opening of multiple isotropic SC gaps in the OD region.

The FS dependence of the SC gap in the OPD K0.4 sample has been reported to be basically consistent with the gap function derived from short-range pairing possibly mediated by the antiferromagnetic (AF) fluctuations, $\Delta(k) = \Delta_1cosk_xcosk_y+\Delta_2(cosk_x+cosk_y)cosk_z/2$ \cite{Yiming2}. It is intriguing to check its validity in the OD region, since the $k_{\rm F}$ position and the SC gap size for each FS are markedly different from those in the OPD region. At first, we neglect the second term and compare the experimental results with the simple formula $\Delta(k) = \Delta_0cosk_xcosk_y$, which is known as one of the extended $s$-wave symmetry \cite{Seo}, as discussed earlier for the K0.4 sample \cite{Hasan1, Nakayama}. This formula predicts the opening of a larger (smaller) gap on a smaller (larger) FS and a sign change of the SC gap between hole and electron FSs, resulting in $s_{\pm}$-wave state. The fitting result ($\Delta_0$ = 9.0 meV) [black dashed line in Fig. 3(d)] shows a basic agreement with the present data, suggesting a dominant contribution of the $s_{\pm}$-wave pairing to the superconductivity. It is also found that the gap size of the $\gamma$ FS is slightly overestimated. This is due to the fact that the $\gamma$ FS is smaller than the $\alpha$ FS, while their gap sizes are comparable. When we introduce the second term, we see a better agreement with the experiment, as indicated by purple dots in Fig. 3(d). Here we use $cosk_z$ = 1, thus $\Delta(k) = \Delta_1cosk_xcosk_y+\Delta_2(cosk_x+cosk_y)/2$, because the present data reflect the electronic structure at $k_z$ $\sim$ 0 \cite{Yiming2, Feng}. For a finite positive $\Delta_2$, $\mid$$\Delta$$\mid$ at the $\Gamma$ point becomes larger than that at the M point, since $cosk_x+cosk_y$ is 2 and 0 at the $\Gamma(0, 0)$ and M$(\pi, 0)$ points, respectively. Therefore, two branches naturally appear in Fig. 3(d) with a single set of fitting parameters ($\Delta_1$, $\Delta_2$) = (8.3 meV, 0.7 meV). The upper and lower branches correspond to $\mid$$\Delta$$\mid$ on the $\Gamma$- and M-centered FSs, respectively. The present results show that the observed multiple nodeless gaps in the OD sample are fitted by a single gap function consistently with the $s_{\pm}$-wave pairing.

\begin{figure}[!t]
\begin{center}
\includegraphics[width=3in]{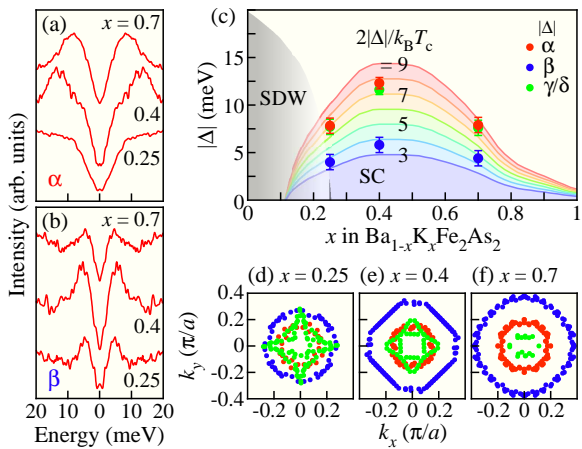}
\end{center}
\caption{
(color online) (a) and (b) Representative symmetrized EDCs in the SC state measured at $k_{\rm F}$ points of the (a) $\alpha$ and (b) $\beta$ FSs for the K0.7 (OD; $T_c$ = 22 K), K0.4 (OPD; $T_c$ = 37 K) and K0.25 (UD; $T_c$ = 26 K) samples. (c) Doping dependence of the averaged SC gap size on each FS. Colored curves show the SC gap size obtained by assuming constant 2$\mid$$\Delta$$\mid$$/k_BT_c$ values (integers from 3 to 9).  (d)-(f) Experimentally determined $k_{\rm F}$ points of the $\alpha$, $\beta$ and $\gamma$/$\delta$ bands [red (gray), blue (dark gray) and green (light gray) circles, respectively]. The $k_{\rm F}$ points of the $\gamma$/$\delta$ FSs are shifted by $Q = (-\pi, 0)$.
}
\end{figure}

We compare in Fig. 4 the present results in the OD sample with those in the UD and OPD samples and discuss the generic properties over a wide doping range. There are three important aspects in the doping dependence of ARPES data. First, the opening of multiple SC gaps is generic among the K0.25, K0.4 and K0.7 samples [Figs. 4(a)-(c)]. While the gap magnitude itself is strongly doping dependent, the obtained $2$$\mid$$\Delta$$\mid$$/k_BT_c$ ratios remain nearly a constant at $k_z$ $\sim$ 0 [$\sim$7.5 and $\sim$4 for larger (for $\alpha$ and $\gamma$) and smaller (for $\beta$) gaps, respectively] over a wide doping range within the present experimental uncertainty [Fig. 4(c)], suggesting that $T_c$ is essentially controlled by the SC gap magnitude (pairing strength).  The linear scaling behavior of $T_c$ and the gap size has been first pointed out in the UD side \cite{Yiming1}. The present results further demonstrate its applicability to the wide doping region in the SC dome including the OD side. Second, as we discussed above, the SC gap in the K0.7 sample is nearly isotropic and nodeless on all FSs. These characteristics are very similar to those observed in the K0.25 and K0.4 samples \cite{Hong, Nakayama, Yiming1}, suggesting the universality of the SC gap symmetry for 0.25  $\leq$ $x$ $\leq$ 0.7. Third, while carrier doping causes drastic changes of the FS areas and the gap values as shown in Figs. 4(c)-(f), the observed multi gaps at each doping level can be commonly fitted by a single gap function assuming short-range pairing with the $s_{\pm}$-wave symmetry [Fig. 3(d) and refs. 4, 6, 7, and 23].

The basic agreement of the FS-dependent gap values with a single gap function is not accidental but rather an intrinsic property of Ba$_{1-x}$K$_x$Fe$_2$As$_2$, and suggests that the nodeless $s_{\pm}$-wave pairing mediated by the short-range AF fluctuations is a highly possible SC mechanism. For the emergence of the nodeless $s_{\pm}$-wave pairing, the interband scattering between hole and electron FS pockets connected by the AF wave vector has been discussed to be an important ingredient \cite{Mazin, Kuroki1, Wang, Seo, Yao, Cvetkovic}. In the K0.25 and K0.4 samples, the inner hole ($\alpha$) and the electron ($\gamma$/$\delta$) FSs are relatively well connected by the shift of $Q = (\pi, 0)$ AF wave vector \cite{Hong, Yiming1} as seen from Figs. 4(d) and 4(e). However, in the K0.7 sample, no obvious overlap of FSs by the shift of $Q = (\pi, 0)$ at $k_z$ $\sim$ 0 is observed [see Fig. 4(f)], suggesting that the FS overlapping is not strictly required. It is known that both the $cosk_xcosk_y$ and $sink_xsink_y$ symmetries can be induced by the next-nearest-neighbor AF fluctuations which are strong in the Fe-based superconductors, and the existence of both FS pockets around $(0, 0)$ and $(\pi, 0)$ would stabilize the $cosk_xcosk_y$ gap symmetry which has the maximum gap amplitude at both the $(0, 0)$ and $(\pi, 0)$ points. The similar case has been made for the high-$T_c$ cuprates where the diamond-like FS stabilizes the $cosk_x$$-$$cosk_y$ gap symmetry induced by the nearest-neighbor AF fluctuations.

In summary, we reported ARPES results on OD Ba$_{1-x}$K$_x$Fe$_2$As$_2$ and compared them with the results on UD and OPD compounds. We observed the opening of multiple nodeless SC gaps in the OD sample, similarly to the results in UD and OPD samples. This result suggests the universality of the pairing symmetry over a wide doping range (0.25 $\leq$ $x$ $\leq$ 0.7). We also found that the FS-dependent gaps can be fitted by a single gap function, $\Delta(k) = \Delta_1cosk_xcosk_y+\Delta_2(cosk_x+cosk_y)cosk_z/2$, consistently with short-range pairing with the $s_{\pm}$-wave symmetry.

We thank X. Dai, Z. Fang and Z. Wang for their valuable discussions and suggestions. We are also grateful to H. Luo, H. H. Wen, J. L. Luo, and N. L. Wang for providing high-quality single crystals of UD and OPD Ba$_{1-x}$K$_x$Fe$_2$As$_2$. This work was supported by grants from JSPS, TRiP-JST, CREST-JST, MEXT of Japan, CAS, NSFC, MOST of China, and NSF of US.

\end{document}